\documentclass[11pt]{amsart}
\usepackage{amsmath}
\usepackage{amssymb}
\usepackage{amscd}
\usepackage{epsfig}
\usepackage{amsxtra}

\newcommand{\bee}{\begin{equation}}
\newcommand{\eee}{\end{equation}}

\newcommand{\be}{\begin{eqnarray}}
\newcommand{\ee}{\end{eqnarray}}
\newcommand{\supp}{\mbox{\rm supp}}

\newcommand{\Vol}{\mbox{\rm Vol}}

\newcommand{\PPD}{\mathcal{PDS}(G)}
\newcommand{\PPDh}{\mathcal{PDS}(\widehat{G})}

\newcommand{\R}{{\mathbb R}}

  \newtheorem{theorem}{Theorem}[section]
  \newtheorem{lemma}[theorem]{Lemma}
  \newtheorem{prop}[theorem]{Proposition}
  \newtheorem{cor}[theorem]{Corollary}
  \newtheorem{fact}[theorem]{Fact}

  \newtheorem{defi}[theorem]{Definition}

\begin{document}
\title[PPDDS measures and Pure Point Diffraction]
{Positive Positive Definite Discrete Strong Almost Periodic Measures and Bragg Diffraction}
\author{Nicolae Strungaru}
\address{Department of Mathematical Sciences, Grant MacEwan University
\\
10700 – 104 Avenue, Edmonton, AB, T5J 4S2;\\
and \\
Institute of Mathematics ``Simon Stoilow'' \\
Bucharest, Romania}
\email{strungarun@macewan.ca}
 \maketitle

\begin{abstract} In this paper we prove that the cone $\PPD$ of positive, positive definite, discrete and strong almost periodic measures has an interesting property: given any positive and positive definite measure $\mu$ smaller than some measure in $\PPD$, then the strong almost periodic part $\mu_S$ of $\mu$ is also in $\PPD$. We then use this result to prove that given a positive weighted comb $\omega$ with finite local complexity and pure point diffraction, any positive comb less than $\omega$ has either trivial Bragg spectrum or a relatively dense set of Bragg peaks.
\end{abstract}

\section{Introduction}

Positive definite measures play an important role in the study of physical diffraction. Given a point set $\Lambda$, its autocorrelation $\gamma$ is a positive definite measure. Thus $\gamma$ is a Fourier transformable measure, and its Fourier transform $\widehat{\gamma}$ is a positive measure, called the diffraction measure. The measure $\widehat{\gamma}$ models the physical diffraction of the structure $\lambda$. Since $\gamma$ is also positive, it follows that both $\gamma$ and $\widehat{\gamma}$ are positive and positive definite measures. This makes the cone of positive and positive definite measures of special interest in the study of diffraction.

The strongest form of long range order is pure point diffraction: a point set $\Lambda$ is pure point diffractive if its diffraction measure $\widehat{\gamma}$ is discrete. This is equivalent to $\gamma$ being a strong almost periodic measure \cite{ARMA}. Moreover, often $\Lambda$ has finite local complexity, and in this case $\gamma$ is also a discrete measure, making it a positive, positive definite, strong almost periodic and discrete. Exactly like positivity and positive definiteness, strong almost periodicity and discreteness are Fourier dual concepts. Hence, given a Delone set  $\Lambda$ with finite local complexity and pure point diffraction, both $\gamma$ and $\widehat{\gamma}$ are positive, positive definite, strong almost periodic and discrete.

The goal of this paper is to introduce and study the cone $\PPD(G)$ of positive, positive definite, discrete and strong almost periodic measures on $G$. This cone of measures already appeared in a hidden way in our earlier work on diffraction. In \cite{NS} we proved that the pure point part $\widehat{\gamma}_{pp}$ of the diffraction of a Meyer set  has a relatively dense support. To obtain this result of interest for the quasicrystal community, we proved the much stronger result that both $\widehat{\gamma}_{pp}$ and $\gamma_{S}$ are discrete measures, and it is easy to see that they both are positive, positive definite and strong almost periodic measures. The ideas from \cite{NS} can be generalized to Theorem \ref{L1}, which is the main result in this paper:

{\bf Theorem \ref{L1}} \rm Let $\nu \in \PPD$ and $0 \leq \mu \leq \nu$ be any positive definite measure. Then $\mu_S \in \PPD$.

The paper is structured as follows:

In Section \ref{S1} we prove Theorem \ref{L1}. We also show in Lemma \ref{L3} and Lemma \ref{L4} that two large classes of measures commonly meet in the mathematics of long range aperiodic order are in $\PPD$.

In Section \ref{S2} we use Theorem \ref{L1} to prove that given a positive Dirac comb $\omega$ with finite local complexity support and pure point spectra, then any non negative measure $\omega'$ less than a constant multiple of $\omega$ has either a relatively dense set of Bragg peaks or no Bragg spectrum. Moreover, we show that if $\omega'$ is not "much smaller" than $\omega$, then we are in the relatively dense Bragg spectra case.

We conclude the paper by taking a short glimpse at the case of real valued weighted Dirac combs.

\section{Preliminaries}

For the entire paper $G$ is a locally compact Abelian group, with dual group $\widehat{G}$. We will denote by $C_C(G)$ the space of compactly supported continuous functions on $G$.

In this Section we will recall the basic notions and results we will need in the paper. All these, with the proofs can be found in \cite{BF} and \cite{ARMA}.

\begin{defi} A measure $\mu$ on $G$ is called {\bf translation bounded} (or {\bf shift bounded}) if for all $f \in C_C(G)$ the function
$f*\mu$ is uniformly continuous and bounded.

We say that $\mu$ is a {\bf positive definite} measure if for all $f \in C_C(G)$ we have $\mu(f*\widetilde{f}) \geq 0$, where $\widetilde{f}(x)=\overline{f(-x)}$.
\end{defi}

Any positive and positive definite measure is translation bounded \cite{BF}.

Argabright and de Lamadrid \cite{ARMA} extended the classical notion of Fourier transformability to measures the following way:

\begin{defi} A measure $\mu$  is called {\bf Fourier transformable} if there exists a measure $\widehat{\mu}$ on $\hat{G}$ so that, for all $f \in C_C(G)$ we have $\check{f} \in L^2(\widehat{\mu})$ and

$$\mu(f*\widetilde{f})= \widehat{\mu}( \left| \check{f} \right|^2 ) \,.$$

In this case $\widehat{\mu}$ is called the {\bf Fourier transform on $\mu$}.
\end{defi}

Bochner's Theorem for continuous positive definite functions can then be extended to positive definite measures:

\begin{theorem}\label{A1}\cite{ARMA} \rm A measure $\mu$ is positive definite if and only if $\mu$ is Fourier transformable and $\widehat{\mu}$ is positive.
\end{theorem}

Thus positive definite measures are automatically Fourier transformable, and their Fourier transform is positive. The other direction is also true, as long as the measure is transformable:

\begin{theorem}\label{A2}\cite{ARMA} \rm Let $\mu$ be a positive measure. If $\mu$ is Fourier transformable, then $\widehat{\mu}$ is positive definite.
\end{theorem}

It follows that positive and positive definite measures are always Fourier transformable, and their Fourier transform is also positive and positive definite. Exactly as in the case of functions, whenever when we can apply the Fourier transform multiple times, the double Fourier transform becomes an involution:

\begin{theorem}\label{A3}\cite{ARMA1} \rm Let $\mu$ be a Fourier transformable measure, so that $\widehat{\mu}$ is also Fourier Transformable. Then
$$\widehat{\widehat{\mu}}= \widetilde{\mu} \,,$$

\noindent where $\widetilde{\mu}(f)=\overline{\mu(\widetilde{f})}$.
\end{theorem}

We conclude the Section by to introducing the notion of strong almost periodic measures and the decomposition ${\mathcal WAP}(G)= {\mathcal SAP}(G) \bigoplus {\mathcal WAP}_0(G)$ of \cite{ARMA}.

\begin{defi} A translation bounded measure $\mu$  is called {\bf strong almost periodic} if for all $f \in C_C(G)$, the function $\mu*f$ is a Bohr-almost periodic function.
\end{defi}

In Section \ref{S4}, we will make use few times of the following simple property of almost periodic measures:

\begin{fact}\cite{NS}  If $\mu$ is strong almost periodic and $\mu\neq 0$, then $\supp(\mu)$ is relatively dense.
\end{fact}

For the exact decomposition ${\mathcal WAP}(G)= {\mathcal SAP}(G) \bigoplus {\mathcal WAP}_0(G)$ we refer the reader to \cite{ARMA}. Since the construction is very technical, we only introduce here a particular case of this decomposition, which is easier to understand and enough for our needs. As the result is not explicitly stated in \cite{ARMA}, but it follows easily for the Theorems in there, we include a proof which the reader might wish to skip.

\begin{theorem}\label{A4} \rm Let $\mu$ be a positive and positive definite measure. Then, there exists a canonical decomposition
\begin{equation}\label{decom}
\mu= \mu_S+ \mu_0 \,,
\end{equation}
so that
 \begin{itemize}
 \item[i)] $\mu_S, \mu_0$ are positive definite,
 \item[ii)] $\mu_S$ is positive and strong almost periodic,
 \item[iii)] $\widehat{\mu_S}= (\widehat{\mu})_{pp} \,;\,  \widehat{\mu_0}= (\widehat{\mu})_{c} \,.$

 \end{itemize}
This decomposition is uniquely determined by these properties. Moreover, $\mu$ is strong almost periodic if and only if $\mu_0=0$.
\end{theorem}
{\sc\bf Proof: \rm  Since $\mu$ is positive and positive definite, it is translation bounded, weakly almost periodic and twice Fourier transformable.

As a weakly almost periodic measure it has an unique decomposition $\mu=\mu_S+\mu_0$ into the strong and null weak almost periodic components. We first show that this decomposition has the desired properties.

Since $\mu$ is positive, we get that $\mu_S$ is positive and strong almost periodic, which proves ii).

We know that $\widehat{\mu}$, and hence $\widehat{\mu}_{pp}$ and $\widehat{\mu}_c$ are positive measures. Since $\widehat{\mu}$ is Fourier transformable, it follows from \cite{ARMA}, Theorem 11.1 that $\widehat{\mu}_{pp}$ and $\widehat{\mu}_c$ are Fourier transformable and

\begin{equation}\label{E1AR} \widehat{\widehat{\mu}_{pp}}=  \widehat{\widehat{\mu}}_S \,;\, \widehat{\widehat{\mu}_{pp}}=  \widehat{\widehat{\mu}}_0 \,.\end{equation}

Moreover, the positivity of $\widehat{\mu}_{pp}$ and $\widehat{\mu}_c$, implies that $\widehat{\widehat{\mu}}_S=\widetilde{\mu}_S$ and $\widehat{\widehat{\mu}}_0=\widetilde{\mu}_0$ are positive definite, and hence $\mu_S$ and $\mu_0$ are also positive definite, which proves i).

To get iii) we proceed as follows: both $\mu$ and $\mu_S$ are positive and positive definite, thus $\mu, \mu_S$ and $\mu_0$ are all twice Fourier transformable. Hence we can apply Theorem \ref{A3} to (\ref{E1AR}), which yields iii).

The uniqueness follows immediately from iii) and the uniqueness of the decomposition $\widehat{\mu}=\widehat{\mu}_{pp}+\widehat{\mu}_c$, while the last claim was proven in \cite{ARMA}.

}\qed

It follows that the class of strong almost periodic measures is important for diffraction:

\begin{cor}\label{A4} Let $\mu$ be a positive and positive definite measure on $G$. Then $\mu$ is strong almost periodic if and only if $\widehat{\mu}$ is discrete.
\end{cor}

As proven in \cite{ARMA}, the decomposition of (\ref{decom}) has a natural extension to the space of translation bounded weakly almost periodic measures \cite{ARMA}, and on this space
$\mu \to \mu_S$ is a linear operator, which is positive and thus preserves inequalities. Thus we get:

\begin{fact} Let $\mu \leq \nu$ be two positive and positive definite measures. Then $\mu_S \leq \nu_S$.
\end{fact}

\section{The cone $\PPD$ }\label{S1}

Throughout this paper we will denote by $\PPD$ the set of all positive, positive definite, strong almost periodic and discrete measures on $G$, that is:

$$\PPD = \{ \mu | \mu \geq 0 \mbox{ is positive definite, strong almost periodic and discrete} \} \,.$$

It is easy to see that this space is closed under addition and multiplication by positive scalars.

We start with a simple Lemma, which is an immediate consequence of Theorem \ref{A1}, Theorem \ref{A2}, Theorem \ref{A3} and Corollary \ref{A4}.

\begin{lemma}\label{L2} \rm If $\mu \in \PPD$ then $\mu$ is Fourier Transformable and $\widehat{\mu} \in \PPDh$. In particular, Fourier transform is a homeomorphism from $\PPD$ to $\PPDh$, both with the vague topology.
\end{lemma}

Let us note that neither $\PPD$ nor $\PPDh$ is closed in the vague topology. If a translation bounded measure $\mu$ is a vague limit of measures in $\PPD$, then $\mu$ is positive and positive definite, but it is not necessarily discrete or strong almost periodic.

Now we are ready to prove the main result in this paper.

\begin{theorem}\label{L1} \rm Let $\nu \in \PPD$ and $0 \leq \mu \leq \nu$ be any positive definite measure. Then $\mu_S \in \PPD$.
\end{theorem}
{\sc\bf Proof: \rm

Since $\mu$ is positive and positive definite, it follows from Corollary \ref{A4} that $\mu_S$ is positive, positive definite and strong almost periodic.

Moreover, $0 \leq \mu \leq \nu$ are positive and positive definite measures, and hence $0 \leq \mu_S \leq \nu_S$. This implies that $\mu_S$ is also discrete. Indeed, since $\nu \in \PPD$, we have $\nu=\nu_S$ is a discrete measure, and using $0 \leq \mu_S \leq \nu$, we get that $\mu_S$ is also discrete. This concludes the proof.

}\qed

By combining Theorem \ref{L1} with Lemma \ref{L2}, we also get:

\begin{cor} \rm Let $\nu \in \PPD$ and $0 \leq \mu \leq \nu$ be any positive definite measure. Then  $\widehat{\mu}_{pp} \in \PPDh$.
\end{cor}

We conclude this Section by introducing a large subclass of $\PPD$, which appeared recently in the study of long range order \cite{BM}, \cite{CR}, \cite{LR}, \cite{BLS}. Since this result is not relevant for the remaining of the paper, we skip the definition of a cut and project scheme, and refer the reader instead to one of these papers.

\begin{lemma}\label{L3} \rm Let $(G \times H, L)$ be a cut and project scheme, and let $h \in C_C(H)$. If $h$ is positive and positive definite, then
$$ \sum_{(x,x^\star) \in L} h(x^\star) \delta_x =: \omega_h \in \PPD \,.$$
\end{lemma}
{\sc\bf Proof: \rm By construction $\omega_h$ is positive and discrete. Also, it is strong almost periodic by \cite{LR}.

The only thing left to prove is the positive definiteness of $\omega_h$. But this follows immediately from the positive definiteness of $h$. Indeed, we get that the function $x \to h(x^\star)$ is a positive definite function on $\pi_1(L)$, thus the measure $\omega_h$ is positive definite as a discrete measure on $\pi_1(L)$ and hence a positive definite measure on $G$ \cite{ARMA}.
}\qed

It is easy to see that the condition $h \in C_C(H)$ can be further weakened to $h$ uniformly continuous, bounded and admissible (see \cite{LR} for the definition).

\section{Diffraction}\label{S2}
\subsection{A review of Diffraction}

In this Section we show that Theorem \ref{L1} can be used to prove that a very large class of weighted Dirac combs have a relatively dense set of Bragg peaks. We start by reviewing the theory of mathematical diffraction. For the entire section  $\{ A_n \}_n$ will denote a fixed but arbitrary van Hove sequence (see for example \cite{NS3} for the definition).

Let $\omega$ be a translation bounded measure on $G$. We define

$$\gamma_n :=\frac{ \omega|_{A_n} * \widetilde{\omega|_{A_n}} }{\Vol(A_n)} \,.$$

It was shown in \cite{BL} that for a translation bounded measure $\omega$, there exists a space ${\mathcal M}_K^C(\R^d)$ which is compact in the vague topology so that $\gamma_n \in {\mathcal M}_K^C(\R^d)$ for all $n$. It follows that the sequence $\gamma_n$ always has cluster points.

\begin{defi}Any cluster point $\gamma$ of the sequence $\gamma_n$ is called an {\bf autocorrelation} of $\omega$.

Any such measure $\gamma$ is positive definite, thus Fourier transformable. Its Fourier transform $\widehat{\gamma}$ is called a {\bf diffraction measure} for $\omega$.
\end{defi}

For a Delone set $\Lambda$, by the autocorrelation of $\Lambda$ we understand the autocorrelation of the measure $\delta_\Lambda:= \sum_{x \in \Lambda} \delta_x \,.$

If $\gamma$ is an autocorrelation of $\omega$, by eventually replacing $\{ A_n \}$ by a subsequence we can always assume that

$$\gamma = \lim_n \gamma_n \,.$$

If we have two translation bounded measures $\omega, \omega'$, and a van Hove sequence $\{ A_n \}$, then $\omega$ has an autocorrelation $\gamma$ with respect of a subsequence $\{ A_{k_n} \}$, and $\omega'$ has an autocorrelation $\gamma'$ with respect to a subsequence of $\{ A_{k_n} \}$. Thus we get:

\begin{fact} Given two translation bounded measures $\omega, \omega'$, and a van Hove sequence $\{ A_n \}$, there exists a subsequence of $\{ A_n \}$ so that both $\omega$ and $\omega'$ have well defined autocorrelations with respect to this subsequence.
\end{fact}

For the remaining of the paper we will adopt the following convention:

\begin{center}
\fbox{\parbox[c]{140mm}
{Given two or more translation bounded measures, their autocorrelations are assumed to be calculated with respect to the same van Hove sequence.
}
}
\end{center}

One notion which will play an important role in the rest of the paper is the notion of Finite Local Complexity. We say that a point set $\Lambda$ has {\bf Finite Local Complexity} if for all compact sets $K \subset G$, the set $\Delta \cap K$ is finite, where $\Delta := \{ x-y |x,y \in \Lambda \}$.

It is easy to see that the autocorrelation measures have the following properties:

\begin{prop} \rm Let $\omega$ be a translation bounded measure, and let $\gamma$ be an autocorrelation of $\omega$. Then
\begin{itemize}
\item[i)] $\gamma$ is positive definite.
\item[ii)] If $\omega$ is positive, then $\gamma$ is positive.
\item[iii)] If $\omega= \sum_{x \in \Lambda} \omega_x \delta_x$ and $\Lambda$ has finite local complexity, then $\supp(\gamma) \subset \Lambda- \Lambda$. In particular, $\gamma$ is discrete.
\end{itemize}
\end{prop}

We now prove a simple Lemma, which we will use few times in the rest of the paper.

\begin{lemma}\label{L4} If $0 \leq \omega' \leq \omega$ and $\gamma', \gamma$ are autocorrelations of $\omega'$ respectively $\omega$, then $\gamma' \leq \gamma$.
\end{lemma}

{\sc\bf Proof: \rm Let $\{ A_n \}$ be the van Hove sequence with respect to which the autocorrelations $\gamma', \gamma$ are calculated. As usual, we denote by $\omega'_n$ and $\omega_n$ the restrictions of $\omega'$ and $\omega$ to $A_n$. Then, as convolutions of positive measures, we have

$$\omega_n*(\widetilde{\omega_n-\omega'_n}) \geq 0 \,;\, \widetilde{\omega'_n}*(\omega_n-\omega'_n) \geq 0 \,.$$

Thus

$$\gamma_n \geq \omega_n*\widetilde{\omega'_n} \geq \gamma'_n \,.$$

Since any vague limit of positive measures is positive, we get that $\gamma-\gamma' = \lim_n (\gamma_n -\gamma'_n) \geq 0 \,.$

}\qed

We complete this subsection by observing that for the class positive translation bounded combs with finite local complexity support, there is a strong connection between pure point diffraction and the $\PPD$ cone:

\begin{lemma}\label{L4} \rm Let $\omega $ be positive translation bounded comb so that $\supp(\omega)$ has finite local complexity. Then $\omega$ is pure point diffractive if and only if $\gamma \in \PPD$.
\end{lemma}

{\sc\bf Proof: \rm We know that $\gamma$ is positive and positive definite and $\gamma$ is discrete. It follows that $\gamma \in \PPD$ if and only if $\gamma$ is strong almost periodic, if and only if $\widehat{\gamma}$ is discrete.
}\qed

In particular we get:

\begin{cor} \rm Let $\Lambda $ be a Delone set with finite local complexity. Then $\Lambda$ is pure point diffractive if and only if $\gamma \in \PPD$.
\end{cor}

Recently it was proven by Lenz and I that an interesting class of measures satisfy the conditions of Lemma \ref{L4}.

\begin{lemma}\cite{LS2} \rm Let $\omega$ be a weakly almost periodic measure. Then $\omega$ has an unique autocorrelation $\gamma$ and $\widehat{\gamma}$ is discrete.
\end{lemma}

An immediate consequence of this is:

\begin{cor} \rm Let $\omega$ be a positive weakly almost periodic measure. If $\sup(\omega)$ has finite local complexity, then $\gamma \in \PPD$.
\end{cor}

\subsection{Measures with a relative dense set of Bragg peaks}\label{S4}

In this subsection we show that Theorem \ref{L1} can be used to prove that a very large class of weighted Dirac combs have a relatively dense set of Bragg peaks. The main result in this subsection is
Theorem \ref{T1} below.

\begin{theorem}\label{T1} \rm Let $\omega:= \sum_{x \in \Lambda} \omega_x \delta_x \, ; \, \omega':= \sum_{x \in \Lambda} \omega'_x \delta_x$ be two positive translation bounded measures, so that $0 \leq \omega' \leq C \omega \,,$ for some $C \geq 0$. Let $\gamma, \gamma'$ be autocorrelations of $\omega$ respectively $\omega'$.

If $\Lambda$ has finite local complexity, $\gamma \neq 0$ and $\omega$ is pure point diffractive, then
\begin{itemize}
\item[i)] $\widehat{\gamma'}_{pp} \in \PPDh$,
\item[ii)] $\widehat{\gamma'}_{c}$ is strong almost periodic,
\item[iii)] $\supp(\gamma'_S) \subset \Lambda- \Lambda$ and $\supp(\gamma'_0) \subset \Lambda- \Lambda$,
\item[iv)] Each of $\supp(\widehat{\gamma'_{pp}})$ and $\widehat{\gamma'}_c$ is either empty or relatively dense,
\end{itemize}

Moreover, if there exists some $C_1 >0$ so that the set
$$ \Gamma := \{ x \in \Lambda |  C_1 \leq \omega'_x \}$$
is relatively dense, then $\omega'$ has a relatively dense set of Bragg peaks.
\end{theorem}
{\sc\bf Proof: \rm

The basic idea of the proof is simple. Since $0 \leq \omega' \leq C\omega$, we will get that $0 \leq \gamma' \leq \gamma$, and we are exactly in the situation of Theorem \ref{L1}. This implies that $\gamma' \in \PPD$, and that $0 \leq \gamma'_S \leq \gamma$. Then i), ii) iii) and iv) are immediate consequences of these two facts.

For the last claim, if $\Gamma$ relatively dense, we prove that there exists a finite set $F$ so that $\omega \leq \omega' * \delta_F$. This implies that $\gamma \leq \gamma'*\delta_F*\widetilde{\delta_F}$ and it is easy to conclude that $\gamma'_S \neq 0$.

${\bf i)}$: Since $0 \leq \omega' \leq C \omega$ we get that
$$0 \leq \gamma' \leq C^2 \gamma \,.$$

Since $C^2 \gamma \in \PPD$, by  Theorem \ref{L1} we get ${\gamma'}_S \in \PPD$, which proves i).

\vspace{.2cm}
${\bf ii)}$: We have seen in i) that $0 \leq \gamma' \leq C^2 \gamma$, thus $\gamma'$ is a discrete measure.

It follows that both $\widehat{\gamma'}$ and $\widehat{\gamma'}_{pp}$ are strong almost periodic, and hence so is their difference $\widehat{\gamma}_c$.

\vspace{.2cm}
${\bf iii)}$:  Using again $0 \leq \gamma' \leq C^2 \gamma$ we get

$$0 \leq \gamma'_S \leq C^2 \gamma_S=C^2 \gamma \,.$$

This shows that $\supp(\gamma'_S) \subseteq \Lambda- \Lambda \,.$ Since $\supp(\gamma') \subseteq \Lambda- \Lambda$, we also get $\supp(\gamma'_0) \subset \Lambda- \Lambda$.

\vspace{.2cm}
${\bf iv)}$: Follows immediately from i) and ii).

To prove the last claim, let $C_1>0$ be such that $\Gamma$ is relatively dense.

Since $\Gamma \subset \Lambda$, $\Gamma$ is relatively dense and $\Lambda$ has finite local complexity, it follows (see \cite{NS} for example) that there exists some finite set $F$ so that

$$\Lambda \subset \Gamma+F \,.$$

Moreover, since $\omega$ is translation bounded, there exists a $C_2$ so that $\omega_x \leq C_2$.

Combining these two, we get:

$$0 \leq \omega \leq C_2 \delta_{\Gamma}*\delta_{F} \,.$$

By the definition of $\Gamma$, we have $\omega' \geq C_1 \delta_\Gamma$, and thus $0 \leq \omega \leq \frac{C_2}{C_1} \omega'*\delta_F \,.$

By taking the autocorrelations, we then get:

$$\gamma \leq \frac{C_2^2}{C_1^2} \left(  \gamma'*\delta_{F}*\widetilde{\delta_F} \right) \,,$$

and hence

$$0 \lneqq \gamma=\gamma_S \leq \frac{C_2^2}{C_1^2} \left( \gamma'_S*\delta_{F}*\widetilde{\delta_F} \right) \,.$$

Then $\gamma'_S \neq 0$, and it follows from $iv)$ that $\supp(\widehat{\gamma_{pp}})$ is relatively dense.

}\qed

In particular, for Delone sets we get:

\begin{cor} \rm Let $\Lambda$ be any Delone set with finite local complexity and let $\Lambda'$ be any subset of $\Lambda$. Let $\gamma'$ be the autocorrelation of $\Lambda'$. If $\Lambda$ is pure point diffractive, then each of the  $\supp(\widehat{\gamma'})_{pp}$ and $\supp(\widehat{\gamma'})_{c}$ is either empty or relatively dense.

In particular, any relatively dense subset of $\Lambda$ shows a relatively dense set of Bragg peaks.
\end{cor}

All these results show that given a positive pure point diffractive comb with finite local complexity, any smaller positive comb, as long as it has some Bragg spectra, has a relatively dense set of Bragg peaks. In the remaining of the paper, we show that in the other direction we also get some Bragg spectra, but maybe less: given a positive pure point diffractive comb with finite local complexity, any bigger translation bounded measure always has infinitely many Bragg peaks.

\begin{prop}\label{P1} \rm Let $\omega$ be a positive translation bounded measure with $0 \neq (\gamma_S)_{pp}$. Let $\omega' \geq \omega$ be any translation bounded measure. Then

$$\lim_n \frac{\widehat{\gamma}_{pp}(B_n)}{| B_n |} >0 \,.$$

In particular, $\widehat{\gamma'}_{pp}$ is infinite and thus $\omega'$ has infinitely many Bragg peaks.

\end{prop}
{\sc\bf Proof: \rm Since $\omega \leq \omega'$ we get $\gamma \leq \gamma'$, and hence $\gamma_S \leq \gamma'_S$.

This shows that $\gamma'_S$ has a point component, and hence $\widehat{\gamma'}_{pp}$ is not a null almost periodic measure.

Because it is positive, it follows \cite{NS}, \cite{Len} that the limit

$$\lim_n \frac{\widehat{\gamma}_{pp}(B_n)}{| B_n |} \,.$$

exists and is non-zero.
}\qed

The following result is an immediate consequences of Proposition \ref{P1}:

\begin{cor} \rm Let $\omega$ be a positive translation bounded comb with $\gamma \neq 0$, and let $\omega' \geq \omega$ be any translation bounded measure. If $\supp(\omega)$ has finite local complexity and $\omega$ is pure point diffractive, then

$$\lim_n \frac{\widehat{\gamma'}_{pp}(B_n)}{| B_n |} >0 \,.$$

In particular, the diffraction pattern of $\omega'$ shows infinitely many Bragg peaks.
\end{cor}

Also, by combining Proposition \ref{P1} with Theorem \ref{T1} we get an interesting result: Given a Delone set $\Lambda$ with finite local complexity and pure point diffraction, if we pick any relatively dense subset $\Lambda' \subset \Lambda$ and any measure $\omega$ larger than a constant multiple of $\delta_{\Lambda'}$, then the diffraction pattern of $\omega$ has infinitely many Bragg peaks:

\begin{cor} \rm  Let $\Lambda$ is a pure point diffractive Delone set with finite local complexity, and $\Lambda' \subset \Lambda$ is relatively dense subset of $\Lambda$ and  $\omega$ is a positive translation bounded measure so that $\omega \geq C \delta_{\Lambda'}$ for some $C >0$, then the diffraction patter of $\omega$ has infinitely many Bragg peaks.
\end{cor}

\subsection{Real Valued Combs}

We conclude by extending part of Theorem \ref{T1} to the case of real valued combs.

\begin{theorem} \rm Let $\omega:= \sum_{x \in \Lambda} \omega_x \delta_x \, ; \, \omega':= \sum_{x \in \Lambda} \omega'_x \delta_x$ be two translation bounded measures, so that $\omega \geq 0$ and $-C \omega \leq \omega' \leq C \omega \,,$ for some $C \geq 0$. Let $\gamma, \gamma'$ be autocorrelations of $\omega$ respectively $\omega'$. If $\gamma \neq 0$, $\omega$ is pure point diffractive and $\gamma'$ is twice Fourier transformable\footnote{We need this requirement since we need to apply the results of \cite{ARMA} to $\widehat{\gamma'}$, and since $\omega'$ might not be positive, we are not guaranteed that $\widehat{\gamma'}$ is positive definite.}, then each of $\supp(\widehat{\gamma'}_{pp})$ and $\widehat{\gamma'}_c$ is either empty or relatively dense.

\end{theorem}
{\sc\bf Proof: \rm  We prove that

$$-C^2 \gamma \leq \gamma' \leq C^2 \gamma \,,$$

\noindent in a similar way as in the proof of Lemma \ref{L4}.

Let $\{ A_n \}$ be the van Hove sequence with respect to which the two autocorrelations are computed, and let again $\omega_n, \omega'_n$ denote the restriction of $\omega$ respectively $\omega'$ to $A_n$. Since $C\omega-\omega' \geq 0$ and $C\omega+\omega' \geq 0$, we have

$$0 \leq (C\omega_n -\omega'_n)* \widetilde{(C\omega_n + \omega'_n)}\, \mbox{and} \,  0 \leq (C\omega_n +\omega'_n)* \widetilde{(C\omega_n - \omega'_n)} \,.$$

By expanding and adding these two relations, we get $C^2\gamma_n \geq \gamma'_n$.

Also, by expanding and adding

$$0 \leq (C\omega_n +\omega'_n)* \widetilde{(C\omega_n + \omega'_n)} \, \mbox{and} \,  \leq (C\omega_n -\omega'_n)* \widetilde{(C\omega_n - \omega'_n)}  \,,$$

we get

$$\gamma'_n \geq -C^2 \gamma_n \,.$$

Thus

$$-C^2 \gamma_n \leq \gamma'_n \leq C^2 \gamma_n \,.$$

By taking the vague limit we get that

$$-C^2 \gamma \leq \gamma' \leq C^2 \gamma \,.$$

We now repeat the idea of Theorem \ref{T1}: by taking the projection on the strong almost periodic component, we get
$-C^2 \gamma \leq \gamma'_S \leq C^2 \gamma$ which implies that $\gamma'_S$ is a discrete measure.

Thus, since $\gamma'$ is twice Fourier transformable, $\widehat{\gamma'}_{pp}$ is strong almost periodic. Moreover, $\widehat{\gamma'}$ is also strong almost periodic, and hence so is $\widehat{\gamma'}_c$. This completes the proof.

}\qed

{\large \sc \bf Acknowledgment} I wish to thank Daniel Lenz for his
suggestions and insightful discussions during the preparation of this paper.
I would also like to thank the Faculty of Mathematics, Friedrich-Schiller-Universit$\ddot{a}$t Jena,
for their kind hospitality.

\end{document}